\documentclass[letter]{aa} 

\usepackage{graphicx}
\usepackage{txfonts}
\usepackage{color}
\usepackage{xspace}
\usepackage{siunitx}
\usepackage{amsmath}
\usepackage{amssymb}
\usepackage{xcolor}
\usepackage{dashbox}
\usepackage{framed}
\usepackage{lipsum}
\usepackage{placeins}
\usepackage{stfloats}

\usepackage{hyperref}

\hypersetup{
        colorlinks=true,                                                
        breaklinks=true,
        linkcolor=blue,                                                     
        citecolor=blue,                                                         
        filecolor=blue,                                                     
        urlcolor=blue,
        unicode=false,                                                      
        pdftoolbar=true,                                                    
        pdfmenubar=true,                                                    
        pdffitwindow=false,                                                 
        pdfstartview={Fit},                                                 
        pdftitle={Extended stellar systems in the solar neighborhood III},  
        pdfauthor={Verena F\"urnkranz},                        
        pdfsubject={},                                          
        pdfcreator={Verena F\"urnkranz},                       
        pdfkeywords={}, 
        pdfnewwindow=true,                                                  
        pdfdisplaydoctitle=true                                     
}

\makeatletter
\renewcommand*\aa@pageof{, page \thepage{} of \pageref*{LastPage}}
\makeatother

\definecolor{stefan}{rgb}{0.86, 0.08, 0.24}

\definecolor{joao}{rgb}{0.01, 0.75, 0.24}

\definecolor{verena}{rgb}{0.5, 0.0, 0.5}

\newcommand{\gaia}{\textit{Gaia}\xspace}
\newcommand{\mel}{Mel~111\xspace}

\DeclareSIUnit\year{yr}

\begin{document}

\defcitealias{meingast18}{Paper~I}
\defcitealias{meingast19}{Paper~II}

\title{Extended stellar systems in the solar neighborhood}
\subtitle{III. Like ships in the night: the  Coma Berenices neighbor moving group\thanks{Full Table~\ref{tab:sources} is only available at the CDS via anonymous ftp.}}

\author{Verena F\"urnkranz\inst{1}
        \and Stefan Meingast\inst{1}
        \and Jo\~ao Alves\inst{1,2,3}
        }
            
\institute{Department of Astrophysics, University of Vienna, T\"urkenschanzstrasse 17, 1180 Wien, Austria
\\ \email{verena.fuernkranz@univie.ac.at}
\and
Radcliffe Institute for Advanced Study, Harvard University, 10 Garden Street, Cambridge, MA 02138, USA
\and
Data Science @ Uni Vienna, Faculty of Earth Sciences Geography and Astronomy, University of Vienna, Austria
}

\date{Received February 18, 2019 / accepted March 12, 2019}

\abstract{We report the discovery of a kinematically cold group of stars, located in the immediate neighborhood of the well-known star cluster Coma Berenices (\mel). The new group identified in tangential velocity space as measured by \gaia contains at least 177 coeval members distributed in two subgroups, and appears as a flattened structure parallel to the plane, stretching for about \SI{50}{pc}. More remarkably, the new group, which appears to have formed about \SI{300}{Myr} later than \mel in a different part of the Galaxy, will share essentially the same volume with the older cluster when the centers of both groups will be at their closest in \SI{13}{Myr}. This will result in the mixing of two unrelated populations with different metallicities. The phase of cohabitation for these two groups is about 20-\SI{30}{Myr}, after which the two populations will drift apart. We estimate that temporal cohabitation of such populations is not a rare event in the disk of the Milky Way, and of the order of once per Galactic revolution. Our study also unveils the tidal tails of the \mel cluster.}

\keywords{Stars: kinematics and dynamics -- solar neighborhood -- open clusters and associations: individual: Coma Berenices}

\maketitle

\section{Introduction}
\label{sec:introduction}

Stellar clusters are unique probes of the physical and chemical conditions at their time and place of birth in the Galaxy.  \gaia provides reliable distances and kinematics to a large number of cluster members and with an unprecedented accuracy. This is causing a renewed interest in the field, in particular in validating ideas for which observational data was lacking. For example, long-suspected dynamical features such as tidal tails have now been identified for the nearest cluster to Earth, the Hyades cluster \citep[][hereinafter Paper I]{roser,meingast18}. At the same time, the expected counterparts of old disk clusters and associations are now beginning to be unveiled \citep[][hereinafter Paper II]{ibata,meingast19}. This newly available parameter space promises to open a new window on cluster disruption, the build up of the field population, the quantification of anisotropies in the mass distribution of the Milky Way disk, and the homogenization of different stellar populations.  

Nevertheless, there is room for surprises. In this Letter we present our follow-up work on the ``Extended stellar systems in the solar neighborhood'' series. While a final catalog is in preparation (Meingast et al. 2019), we report here evidence for temporal cohabitation of different stellar populations in the same Galactic volume. The clusters in question, Coma Berenices (\mel) and a newly found moving group in the velocity and spatial neighborhood, are not massive enough for capture of populations to occur, but they will appear in the near future, and for a limited  time, as a multi-population cluster.

\section{Data description and member selection}
\label{sec:data}

\begin{figure*}
        \centering
        \resizebox{1.0\hsize}{!}{\includegraphics[]{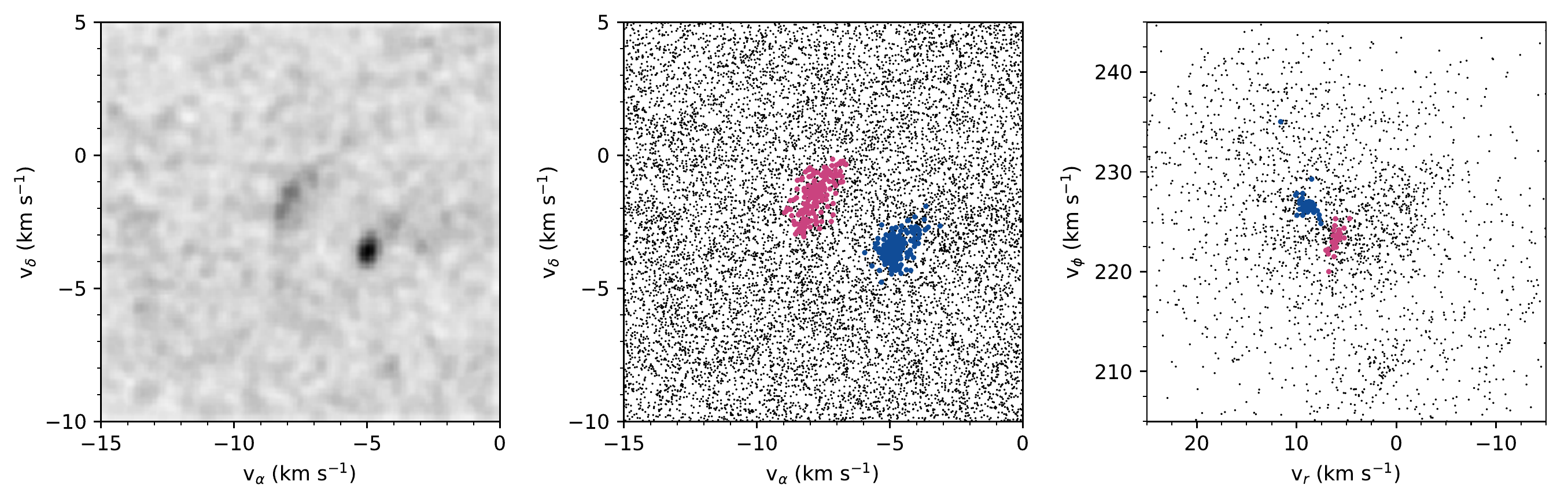}}
        \caption[]{The left panel shows a KDE, using  an Epanechnikov kernel with a bandwidth of \SI{0.4}{\km \per \second} in the tangential velocity space for the \SI{11294}{} stars in our final database. Two prominent overdensities are visible, corresponding to \mel and the new group which are displayed in the middle panel in blue and magenta, respectively. The right panel displays our selection in the v$_{r}$v$_{\phi}$-velocity plane. Small black dots represent all sources from the filtered \textit{Gaia} database.}
    \label{img:pm}
\end{figure*}

As in \citetalias{meingast19}, we detected overdensities in velocity space, given by \gaia DR2 \citep[][]{gaia_mission, gaia_dr2}, with a wavelet decomposition in Galactocentric Cylindrical coordinates\footnote{For details on the coordinate system definition see \citetalias{meingast18}.}. Among the extracted significant peaks, we found the velocity coordinates of the well-known star cluster \mel at $(v_r, v_{\phi}, v_z) =(8.83, 226.93, 6.54)$ \SI{}{\km \per \second}, as well as a nearby second overdensity at $(v_r, v_{\phi}, v_z) =(6.21, 223.57, 5.41)$ \SI{}{\km \per \second}, belonging to a previously unknown stellar population. An extraction of all sources within a \SI{5}{\km \per \second} radius around the identified peaks indicated that both populations do not only share very similar velocities, but are also adjacent in spatial coordinates, making them an interesting case for further investigation on possible cluster interactions.

In order to minimize the error budget, we adopted filtering criteria similar to \citetalias{meingast19}: $\sigma_{\mu_{\alpha,\delta}} / \mu_{\alpha,\delta} < 0.5$, $\sigma_{\varpi} / \varpi < 0.5$, and max$_{\sigma 5D} < 0.5$. Since \mel and its newly discovered neighbor are located well above the Galactic plane, we also restricted our database to \SI{0}{pc} < Z < \SI{150}{pc}, \SI{-75}{pc} < X < \SI{75}{pc}, and \SI{-50}{pc} < Y < \SI{100}{pc}. 

In contrast to the prominent stream identified in \citetalias{meingast19}, here we find structures with smaller spatial extent. In such cases projection effects are minimized and consequently we based our member selection on 2D tangential velocity space (v$_{\alpha}$, v$_{\delta}$) rather than on 3D velocities. We obtained tangential velocities for every source from proper motion and distance data, and applied a further restriction of \SI{-15}{\km \per \second} < v$_{\alpha}$ < \SI{0}{\km \per \second} and \SI{-10}{\km \per \second} < v$_{\delta}$ < \SI{5}{\km \per \second}. After applying these filters, a total number of \SI{11294}{} sources remained. The tangential velocity distribution of the remaining sources is illustrated with a kernel density map in the left panel of Fig.~\ref{img:pm}. In this view, two local overdensities become clearly visible. The tight, point-like structure at (v$_{\alpha}$,v$_{\delta}$) $\sim$ (\mbox{-5, -4) \SI{}{\km \per \second}} contains sources associated with \mel, whereas the elongated arc-shape at (v$_{\alpha}$,v$_{\delta}$) $\sim$ (\mbox{-8, -2) \SI{}{\km \per \second}} represents a previously unknown stellar group.

Following the setup outlined above, we then extracted clustered sources with the density-based algorithm DBSCAN \citep{dbscan}. Specifically for our selection, we manually chose $\mathrm{minPts}=70$, $\epsilon=\;$\SI{0.5}{\km \per \second} for the DBSCAN setup which resulted in two tangential velocity clusters associated with the two apparent overdensities in the left panel of Fig.~\ref{img:pm}. This selection extracted 245 sources of \mel, and 237 stars associated with the new group. Following the previous papers, we additionally restricted the selection by applying a spatial density filter. We tested several setups, where our final criterion excludes all sources that have less than 30 neighbors within \SI{20}{pc}. Finally, we manually removed one star that was located below the main sequence and also showed a large photometric excess factor, indicating contaminated \gaia photometry. This resulted in a final selection of 214 \mel sources and 177 sources for the new group. Table~\ref{tab:table} lists several parameters measured for these two groups.

The middle panel of Fig.~\ref{img:pm} shows the final selection for both groups in the tangential velocity space. While the blue points represent our member selection for \mel, the new group is illustrated in magenta. For clarity, these colors are the same for all the figures presented here. The right panel of Fig.~\ref{img:pm} displays the distribution of the stars in the v$_{r}$v$_{\phi}$-velocity plane, where the two significant overdensities are colored corresponding to our member selection of \mel and the new group. It shows both populations tightly clustered, thus verifying our selection process.

We estimated the contamination level with two methods. First, we applied the same method as described in \citetalias{meingast19}, which extracts sources in a symmetric phase-space region on the opposite side of the Galactic plane. Following the same steps (with adapted measurements), we find a fractional contamination level of only a few percent. Secondly, the velocity distribution in Fig.~\ref{img:pm} reveals that each group contains a few stars which do not fit the general velocity profile of the groups. This closely matches the galactic field contamination estimate outlined above. Here, we chose not to remove these outliers in velocity space from our selection, since such a restriction could only be consistently applied to stars with radial-velocity measurements and not to all sources.

\section{Results and discussion}
\label{sec:discussion}

\subsection{Structure}
\label{sec:structure}

\begin{figure*}
        \centering
        \resizebox{1.0\hsize}{!}{\includegraphics[]{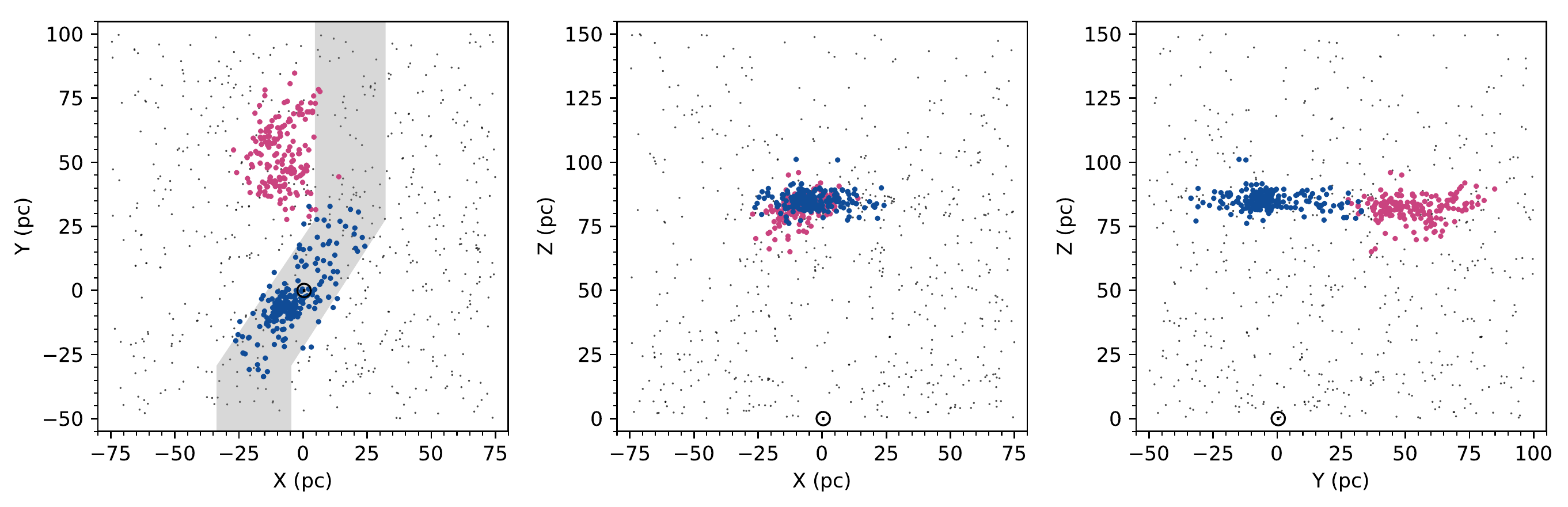}}
        \caption[]{Positions of the final member selection in Galactic Cartesian coordinates. The position of the Sun is indicated with the black circular symbol. The gray shaded area represents the approximate shape of the
tidal tails of  \mel \citep{chumak}. The small black dots correspond to all sources that where identified in our proper-motion clustering application but did not pass the spatial filtering.}
    \label{img:zoom}
\end{figure*}

Figure~\ref{img:zoom} illustrates the final member selection in Galactic Cartesian coordinates. The same distribution projected on the sky is shown in Fig.~\ref{img:proj}. \mel is located at a distance of approximately \SI{85}{pc} from the Sun in the direction of the north Galactic pole. The new group is located at almost the same distance to the Galactic plane and at a similar Galactocentric radius, but about \SI{60}{pc} ahead in the direction of Galactic rotation. 

The \mel selection reveals a flattened shape parallel to the Galactic plane, as well as a pronounced core in the cluster center. The XY distribution of the cluster shows a tilted ellipsoidal structure, with a length of about \SI{60}{pc} and a thickness of about \SI{25}{pc}. Following the discoveries of tidal tails associated with the Hyades \citep{roser, meingast18}, we also compared our findings for \mel to the predicted tail structure as given by \citet{chumak}. The approximate shape of the tails is shown in the left panel of Fig.~\ref{img:zoom}, which is in excellent agreement with our selected \mel sources.

In contrast to \mel, the new group shows different morphological characteristics. Most importantly, it does not have a similarly pronounced core, which is  likely the reason why it has not yet been found. Moreover, the top-down view of the new group members reveals an inhomogeneous distribution of sources, which are arranged in two parallel lanes. The two subgroups show a systematic offset in proper motions, but we do not find a significant difference in space velocities and other physical parameters. Therefore, we argue that the difference in proper motions only results from projection effects. 

We derive stellar masses similar to the previous entries in this paper series by interpolating isochrones for the systems (Sect.~\ref{sec:age}). Figure~\ref{img:mass} in the Appendix shows the resulting present-day mass functions compared to a series of initial mass functions \citep[IMF;][]{kroupa01}, which we used to estimate the birth masses of the systems. In general, we find a good match between the mass function for \mel and the new group, suggesting similar current masses (affected by two-body relaxation and tidal forces). Also, we find that the present-day mass function is in overall good agreement with a \SI{200}{M_\odot} IMF for both systems (especially near the higher-mass end of our selection). However, the measured present-day masses of magnitude-limited samples are generally affected by incompleteness. \citetalias{meingast18} determined the Hyades selection in this mass range to be incomplete by a factor of approximately two. Following this result, but considering the larger distances to the two groups discussed in this manuscript, we estimated the bias to result in a factor of approximately three or more, shifting the mass function closer to a \SI{500}{M_\odot} initial mass estimate. This estimate should be seen as a lower limit, because we did not consider mass loss caused by stellar evolution and tides. Moreover, in Sect.~\ref{sec:age} we show that the new group is most likely several hundred million years younger than \mel. Therefore, if the initial masses of the systems were similar, the current lack of a pronounced core in the new group indicates either 
a different initial condition (cluster vs. association) or a very different dynamical evolution.

\subsection{Age and metallicity}
\label{sec:age}

\begin{figure}
        \centering
        \resizebox{1.0\hsize}{!}{\includegraphics[]{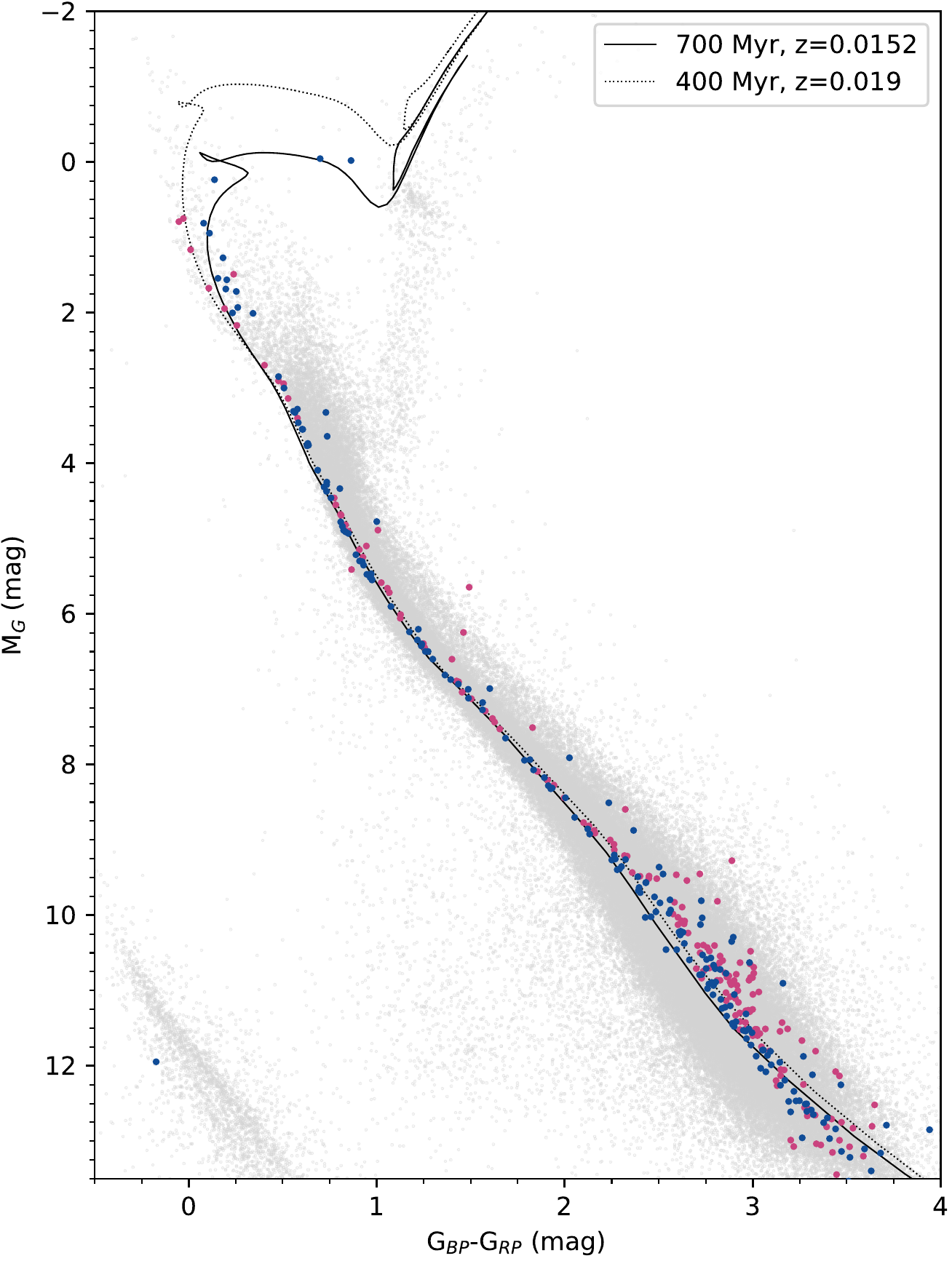}}
        \caption[]{Observational HRD for the member stars of \mel and the new group. The solid line represents the \SI{700}{Myr} PARSEC isochrone with solar metallicity and the dashed line illustrates the 400 Myr PARSEC isochrone with $\mathrm{z} = 0.019$. The gray dots in the background are all sources in our filtered database.}
    \label{img:hrd}
\end{figure}

We present an observational HRD of our member selection in Fig.~\ref{img:hrd}. Both groups show a well-defined main sequence, indicating that each group by itself comprises a coeval stellar population. While \mel hosts two stars beyond the main sequence turn-off, as well as one white dwarf, all selected members of the new group are located on the main sequence. Comparing the two sequences, we find significant differences both near the upper and lower main sequence. The upper main sequence of \mel is located on top of the new group, whereas this offset reverses as we follow the main sequence down to the cooler and less luminous stars, shifting \mel to the bluer part. 

In order to estimate the age of the two groups, we compared our selections with PARSEC isochrones \citep{bressan}. Assuming solar metallicity for \mel \citep{netopil}, the 700 Myr isochrone appears to fit the sequence well. This is also consistent with previously published ages for \mel \citep[e.g.,][]{tang, hrd}. The upper main sequence of the new group indicates a turnoff at higher luminosities compared to \mel and therefore a younger stellar age. Adjusting only the age of the isochrones, we find that a \SI{400}{Myr} isochrone fits well to the upper part of the sequence. However, this adjustment does not match the observed offset near the lower main sequence. 

As also metallicity generally affects the location and shape of the main sequence, we cross-matched our selection with LAMOST DR4 \citep{cui}, resulting in nine matches for \mel and eight matches for the new group. The mean metallicity of the matched sources is $\left[ \mathrm{Fe/H} \right] = -0.117 \pm 0.115$ for \mel and $\left[ \mathrm{Fe/H} \right] = -0.003 \pm 0.093$ for the new group. We note here that the measured metallicity for \mel is not consistent with our previous  assumption of solar metallicity. This difference is likely caused by a systematic offset in the survey and we therefore only take the relative metallicity offset of $\sim$ \SI{0.1}{dex} between both groups into account. The \SI{400}{Myr} isochrone with higher metallicity then also fits well to the lower main sequence of the new group. Clearly, more data and an improved set of models are needed to better age the new group, but the two groups were formed at different times. For the remainder of this Letter we assume their age difference to be about 300 Myr.

\subsection{Kinematics and Galactic orbit}
\label{sec:kinematics}

As pointed out in Sect.~\ref{sec:data}, the two groups have a very similar kinematic profile. An inspection of the Galactocentric Cylindrical velocities of both groups reveals only small differences in the radial and azimuthal velocity component ($\Delta$v$_{r}$ = \SI{2.9}{\km \per \second} and $\Delta$v$_{\phi}$ = \SI{3.5}{\km \per \second}). The vertical velocity component is virtually identical. Interestingly, \mel, lagging behind in the Galactic rotation, is more than \SI{3}{\km \per \second} faster in v$_{\phi}$ compared to the new group. This configuration therefore indicates that \mel could drift into the new group within the next few million years.

To analyze a potential encounter in more detail, we calculated the orbital motions of the groups with \textit{Galpy} \citep{bovy}. We used a predefined axisymmetric setup for the gravitational potential of the Milky Way (MWPotential2014) which comprises a bulge, a disk, and a dark-matter-halo component; it does however not include spiral arms or molecular clouds. For a full description of the orbit, radial-velocity measurements are also required for which we added another quality criterion, limiting the error in radial velocity to $\sigma_{rv}$ <  \SI{2}{\km \per \second}. This was applied to keep the errors of the orbit integrations at a manageable level and at the same time retain the bulk of our selection (about 90\% of the sources which have radial-velocity measurements). Within these limits, we find 61 \mel sources and 26 sources associated with the new group.

We estimated position errors along the integrated orbit by randomly sampling the error distribution in distance, $\mu_{\alpha}$, $\mu_{\delta}$, and radial velocity. Each random sample (total sample size 100) was then integrated independently, resulting in a distribution of positions for each time-step. Here, the error most significantly depends on the radial-velocity measurement. Together with the location of the groups near the Galactic north pole, this results in relatively large errors in the vertical position.

\begin{figure}
        \centering
        \resizebox{1.0\hsize}{!}{\includegraphics[]{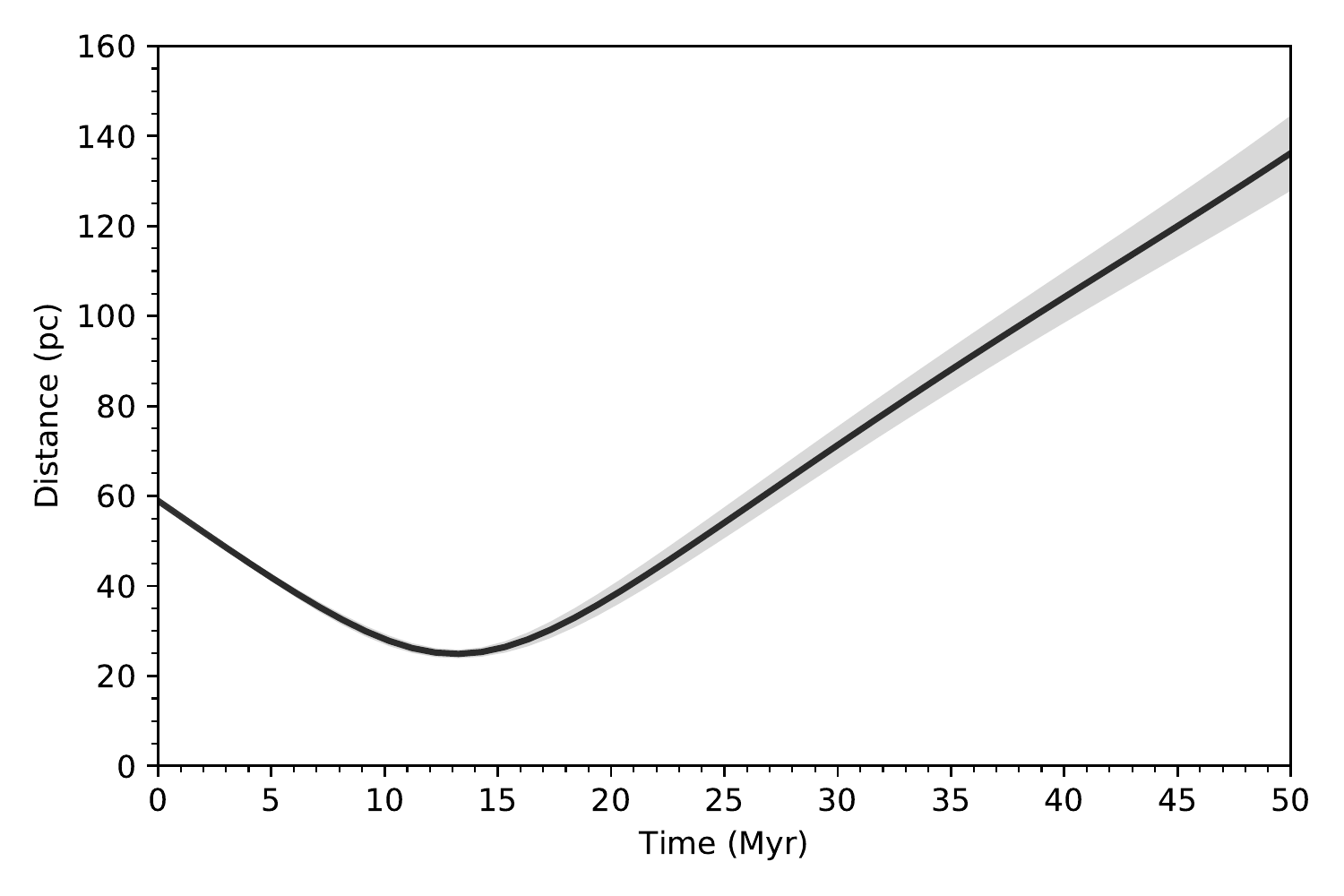}}
        \caption[]{Distance between \mel and the new group as a function of time. The black solid line displays the mean distance and the gray shaded area corresponds to the 3-$\sigma$ error. We find a minimum at $t = $\SI{13}{Myr}.}
    \label{img:dist}
\end{figure}

Figure~\ref{img:dist} shows the distance between the group centers (determined as the mean position of sources) as a function of time going forward along their orbit, including the 3-$\sigma$ error interval. Remarkably, the two groups continue to converge for the next few million years where we find a minimum distance of \SI{25}{pc} in \SI{13}{Myr} after which they start to drift apart again. Here, the escape velocity at a distance of \SI{25}{pc} even for a mass estimate of \SI{500}{M_\odot} for \mel is only \SI{0.4}{\km \per \second}. Thus, the relative velocity offset between the groups is too large for a potential merging event. Nevertheless, given their spatial extent of at least \SI{50}{pc}, the two systems will essentially share the same volume for about 20 to \SI{30}{Myr}. Figure~\ref{img:galpy} shows the position of all sources in Galactic Cartesian coordinates both now and at the time of minimum distance. We also note that the orbits are integrated independently and that we did not add additional gravitational potentials for the individual groups.

This close encounter between two stellar populations encourages speculations on how often such events occur in the Galaxy. To test this, we created a simple setup of open clusters scattered across the entire Galactic disk, integrated their orbits \SI{100}{Myr} forward, and calculated the average number of encounters (distance between two groups < \SI{20}{pc)}. Specifically, we started with estimating the top-down surface density of clusters in the disk. The Webda database \citep{webda} lists a total of 345 open clusters (including loose associations and moving groups) within \SI{1}{kpc}. This number translates into an average of about 55 such objects per \SI{}{kpc \squared} in the Galactic plane. Since the actual distribution of star clusters in the Galaxy is unknown and their radial distribution could even be a function of cluster age \citep[e.g.,][]{scheepmaker}, we favored an isotropic setup for our toy model with a total number of \SI{50000}{} open clusters for a \SI{30}{kpc}-wide disk. Furthermore, we randomly sampled the velocity distribution directly from the \gaia DR2 measurements of all stars in the solar neighborhood. A forward integration for \SI{100}{Myr} revealed an average rate of 200 encounters per million years across the entire disk. Therefore, for our \SI{50000}{} mock clusters, each cluster should have on average one encounter every \SI{250}{Myr}, indicating that such meetings of groups can happen about once per Galactic revolution.

\section{Summary and conclusions}
\label{sec:summary}

Following the previous papers in this series, we used position and velocity data provided by \gaia DR2 to analyze two specific overdensities in velocity space (Fig.~\ref{img:pm}). The first, more prominent peak corresponds to \mel, while the second overdensity, separated by only a few kilometres per second, marks a previously unknown stellar population. Moreover, these groups do not only share similar kinematics (Table~\ref{tab:table}), but they are also currently only about \SI{60}{pc} apart (Fig.~\ref{img:zoom}). The spatial arrangement of the source selection for \mel also shows striking similarity to theoretically predicted tidal tails. The new group however does not show a pronounced core, but instead appears to be arranged in two parallel lanes, which are not clearly separable in velocity space. A comparison of the main sequences of the groups, taking into account metallicity differences, reveals an age of about \SI{700}{Myr} for \mel in agreement with previously obtained results. The age of the newly discovered group appears to be best represented by a \SI{400}{Myr} isochrone (Fig.~\ref{img:hrd}).

We also analyzed the kinematics of the groups in order to investigate a possible future interaction. By integrating individual orbits we find that both groups currently converge, with a minimum distance of only \SI{25}{pc} between the cluster centers \SI{13}{Myr} from now (Fig.~\ref{img:dist}), resulting in temporary mixing of two unrelated stellar populations (Fig.~\ref{img:galpy}) for about 20-\SI{30}{Myr}. The masses of the systems are however not large enough to overcome the velocity difference, preventing a merging process. A toy setup and forward integration of mock open clusters distributed across the entire Galactic disk reveals that such encounters can happen at a rate of about one per Galactic revolution for each cluster. Thus, the observed encounter between \mel and the newly discovered group is probably not a unique phenomenon. This process is reminiscent, although not proof, of the multi-populations found in the massive globular clusters \citep{Bedin2004}.

\begin{acknowledgements}
We wish to thank the anonymous referee for his/her useful comments, which improved both the clarity and quality of this study. This work has made use of data from the European Space Agency (ESA) mission \gaia (\url{https://www.cosmos.esa.int/gaia}), processed by the \gaia Data Processing and Analysis Consortium (DPAC, \url{https://www.cosmos.esa.int/web/gaia/dpac/consortium}). Funding for the DPAC has been provided by national institutions, in particular the institutions participating in the \gaia Multilateral Agreement.
This research made use of Astropy, a community-developed core Python package for Astronomy \citep{astropy}.
This research has made use of "Aladin sky atlas" developed at CDS, Strasbourg Observatory, France \citep{bonnarel00}.
We also acknowledge the various Python packages that were used in the data analysis of this work, including NumPy \citep{numpy}, SciPy \citep{scipy}, scikit-learn \citep{scikit-learn}, scikit-image \citep{scikit-image}, and Matplotlib \citep{matplotlib}.
This research has made use of the SIMBAD database operated at CDS, Strasbourg, France \citep{simbad}.
\end{acknowledgements}

\bibliography{references}

\begin{thebibliography}{29}
\expandafter\ifx\csname natexlab\endcsname\relax\def\natexlab#1{#1}\fi

\bibitem[{{Astropy Collaboration} {et~al.}(2018){Astropy Collaboration},
  {Price-Whelan}, {Sip{\H o}cz}, {G{\"u}nther}, {Lim}, {Crawford}, {Conseil},
  {Shupe}, {Craig}, {Dencheva}, {Ginsburg}, {VanderPlas}, {Bradley},
  {P{\'e}rez-Su{\'a}rez}, {de Val-Borro}, {Aldcroft}, {Cruz}, {Robitaille},
  {Tollerud}, {Ardelean}, {Babej}, {Bach}, {Bachetti}, {Bakanov}, {Bamford},
  {Barentsen}, {Barmby}, {Baumbach}, {Berry}, {Biscani}, {Boquien}, {Bostroem},
  {Bouma}, {Brammer}, {Bray}, {Breytenbach}, {Buddelmeijer}, {Burke},
  {Calderone}, {Cano Rodr{\'{\i}}guez}, {Cara}, {Cardoso}, {Cheedella},
  {Copin}, {Corrales}, {Crichton}, {D'Avella}, {Deil}, {Depagne}, {Dietrich},
  {Donath}, {Droettboom}, {Earl}, {Erben}, {Fabbro}, {Ferreira}, {Finethy},
  {Fox}, {Garrison}, {Gibbons}, {Goldstein}, {Gommers}, {Greco}, {Greenfield},
  {Groener}, {Grollier}, {Hagen}, {Hirst}, {Homeier}, {Horton}, {Hosseinzadeh},
  {Hu}, {Hunkeler}, {Ivezi{\'c}}, {Jain}, {Jenness}, {Kanarek}, {Kendrew},
  {Kern}, {Kerzendorf}, {Khvalko}, {King}, {Kirkby}, {Kulkarni}, {Kumar},
  {Lee}, {Lenz}, {Littlefair}, {Ma}, {Macleod}, {Mastropietro}, {McCully},
  {Montagnac}, {Morris}, {Mueller}, {Mumford}, {Muna}, {Murphy}, {Nelson},
  {Nguyen}, {Ninan}, {N{\"o}the}, {Ogaz}, {Oh}, {Parejko}, {Parley}, {Pascual},
  {Patil}, {Patil}, {Plunkett}, {Prochaska}, {Rastogi}, {Reddy Janga},
  {Sabater}, {Sakurikar}, {Seifert}, {Sherbert}, {Sherwood-Taylor}, {Shih},
  {Sick}, {Silbiger}, {Singanamalla}, {Singer}, {Sladen}, {Sooley},
  {Sornarajah}, {Streicher}, {Teuben}, {Thomas}, {Tremblay}, {Turner},
  {Terr{\'o}n}, {van Kerkwijk}, {de la Vega}, {Watkins}, {Weaver}, {Whitmore},
  {Woillez}, {Zabalza}, \& {Astropy Contributors}}]{astropy}
{Astropy Collaboration}, {Price-Whelan}, A.~M., {Sip{\H o}cz}, B.~M., {et~al.}
  2018, \aj, 156, 123

\bibitem[{{Bedin} {et~al.}(2004){Bedin}, {Piotto}, {Anderson}, {King},
  {Cassisi}, \& {Momany}}]{Bedin2004}
{Bedin}, L.~R., {Piotto}, G., {Anderson}, J., {et~al.} 2004, Memorie della
  Societa Astronomica Italiana Supplementi, 5, 105

\bibitem[{{Bonnarel} {et~al.}(2000){Bonnarel}, {Fernique}, {Bienaym{\'e}},
  {Egret}, {Genova}, {Louys}, {Ochsenbein}, {Wenger}, \&
  {Bartlett}}]{bonnarel00}
{Bonnarel}, F., {Fernique}, P., {Bienaym{\'e}}, O., {et~al.} 2000, \aaps, 143,
  33

\bibitem[{{Bovy}(2015)}]{bovy}
{Bovy}, J. 2015, The Astrophysical Journal Supplement Series, 216, 29

\bibitem[{{Bressan} {et~al.}(2012){Bressan}, {Marigo}, {Girardi}, {Salasnich},
  {Dal Cero}, {Rubele}, \& {Nanni}}]{bressan}
{Bressan}, A., {Marigo}, P., {Girardi}, L., {et~al.} 2012, \mnras, 427, 127

\bibitem[{{Casewell} {et~al.}(2006){Casewell}, {Jameson}, \&
  {Dobbie}}]{casewell}
{Casewell}, S.~L., {Jameson}, R.~F., \& {Dobbie}, P.~D. 2006, \mnras, 365, 447

\bibitem[{{Chumak} \& {Rastorguev}(2006)}]{chumak}
{Chumak}, Y.~O. \& {Rastorguev}, A.~S. 2006, Astronomy Letters, 32, 446

\bibitem[{{Cui} {et~al.}(2012){Cui}, {Zhao}, {Chu}, {Li}, {Li}, {Zhang}, {Su},
  {Yao}, {Wang}, {Xing}, {Li}, {Zhu}, {Wang}, {Gu}, {Luo}, {Xu}, {Zhang},
  {Liu}, {Zhang}, {Yang}, {Cao}, {Chen}, {Chen}, {Chen}, {Chen}, {Chu}, {Feng},
  {Gong}, {Hou}, {Hu}, {Hu}, {Hu}, {Jia}, {Jiang}, {Jiang}, {Jiang}, {Jin},
  {Li}, {Li}, {Li}, {Liu}, {Liu}, {Lu}, {Mao}, {Men}, {Qi}, {Qi}, {Shi},
  {Tang}, {Tao}, {Wang}, {Wang}, {Wang}, {Wang}, {Wang}, {Wang}, {Wang},
  {Wang}, {Wang}, {Wang}, {Wang}, {Wang}, {Xu}, {Xu}, {Yang}, {Yu}, {Yuan},
  {Yuan}, {Zhai}, {Zhang}, {Zhang}, {Zhang}, {Zhao}, {Zhou}, {Zhou}, {Zhu}, \&
  {Zou}}]{cui}
{Cui}, X.-Q., {Zhao}, Y.-H., {Chu}, Y.-Q., {et~al.} 2012, Research in Astronomy
  and Astrophysics, 12, 1197

\bibitem[{Ester {et~al.}(1996)Ester, Kriegel, Sander, \& Xu}]{dbscan}
Ester, M., Kriegel, H.-P., Sander, J., \& Xu, X. 1996, in Proceedings of the
  Second International Conference on Knowledge Discovery and Data Mining,
  KDD'96 (AAAI Press), 226--231

\bibitem[{{Gaia Collaboration} {et~al.}(2018{\natexlab{a}}){Gaia
  Collaboration}, {Babusiaux}, {van Leeuwen}, {Barstow}, {Jordi}, {Vallenari},
  {Bossini}, {Bressan}, {Cantat-Gaudin}, {van Leeuwen}, {Brown}, {Prusti}, {de
  Bruijne}, {Bailer-Jones}, {Biermann}, {Evans}, {Eyer}, {Jansen}, {Klioner},
  {Lammers}, {Lindegren}, {Luri}, {Mignard}, {Panem}, {Pourbaix}, {Randich},
  {Sartoretti}, {Siddiqui}, {Soubiran}, {Walton}, {Arenou}, {Bastian},
  {Cropper}, {Drimmel}, {Katz}, {Lattanzi}, {Bakker}, {Cacciari},
  {Casta{\~n}eda}, {Chaoul}, {Cheek}, {De Angeli}, {Fabricius}, {Guerra},
  {Holl}, {Masana}, {Messineo}, {Mowlavi}, {Nienartowicz}, {Panuzzo},
  {Portell}, {Riello}, {Seabroke}, {Tanga}, {Th{\'e}venin}, {Gracia-Abril},
  {Comoretto}, {Garcia-Reinaldos}, {Teyssier}, {Altmann}, {Andrae}, {Audard},
  {Bellas-Velidis}, {Benson}, {Berthier}, {Blomme}, {Burgess}, {Busso},
  {Carry}, {Cellino}, {Clementini}, {Clotet}, {Creevey}, {Davidson}, {De
  Ridder}, {Delchambre}, {Dell'Oro}, {Ducourant}, {Fern{\'a}ndez-
  Hern{\'a}ndez}, {Fouesneau}, {Fr{\'e}mat}, {Galluccio}, {Garc{\'\i}a-Torres},
  {Gonz{\'a}lez-N{\'u}{\~n}ez}, {Gonz{\'a}lez-Vidal}, {Gosset}, {Guy},
  {Halbwachs}, {Hambly}, {Harrison}, {Hern{\'a}ndez}, {Hestroffer}, {Hodgkin},
  {Hutton}, {Jasniewicz}, {Jean-Antoine-Piccolo}, {Jordan}, {Korn},
  {Krone-Martins}, {Lanzafame}, {Lebzelter}, {L{\"o}ffler}, {Manteiga},
  {Marrese}, {Mart{\'\i}n-Fleitas}, {Moitinho}, {Mora}, {Muinonen}, {Osinde},
  {Pancino}, {Pauwels}, {Petit}, {Recio-Blanco}, {Richards}, {Rimoldini},
  {Robin}, {Sarro}, {Siopis}, {Smith}, {Sozzetti}, {S{\"u}veges}, {Torra}, {van
  Reeven}, {Abbas}, {Abreu Aramburu}, {Accart}, {Aerts}, {Altavilla},
  {{\'A}lvarez}, {Alvarez}, {Alves}, {Anderson}, {Andrei}, {Anglada Varela},
  {Antiche}, {Antoja}, {Arcay}, {Astraatmadja}, {Bach}, {Baker},
  {Balaguer-N{\'u}{\~n}ez}, {Balm}, {Barache}, {Barata}, {Barbato}, {Barblan},
  {Barklem}, {Barrado}, {Barros}, {Bartholom{\'e} Mu{\~n}oz}, {Bassilana},
  {Becciani}, {Bellazzini}, {Berihuete}, {Bertone}, {Bianchi}, {Bienaym{\'e}},
  {Blanco-Cuaresma}, {Boch}, {Boeche}, {Bombrun}, {Borrachero}, {Bouquillon},
  {Bourda}, {Bragaglia}, {Bramante}, {Breddels}, {Brouillet},
  {Br{\"u}semeister}, {Brugaletta}, {Bucciarelli}, {Burlacu}, {Busonero},
  {Butkevich}, {Buzzi}, {Caffau}, {Cancelliere}, {Cannizzaro}, {Carballo},
  {Carlucci}, {Carrasco}, {Casamiquela}, {Castellani}, {Castro-Ginard},
  {Charlot}, {Chemin}, {Chiavassa}, {Cocozza}, {Costigan}, {Cowell}, {Crifo},
  {Crosta}, {Crowley}, {Cuypers}, {Dafonte}, {Damerdji}, {Dapergolas}, {David},
  {David}, {de Laverny}, {De Luise}, {De March}, {de Martino}, {de Souza}, {de
  Torres}, {Debosscher}, {del Pozo}, {Delbo}, {Delgado}, {Delgado}, {Diakite},
  {Diener}, {Distefano}, {Dolding}, {Drazinos}, {Dur{\'a}n}, {Edvardsson},
  {Enke}, {Eriksson}, {Esquej}, {Eynard Bontemps}, {Fabre}, {Fabrizio},
  {Faigler}, {Falc{\~a}o}, {Farr{\`a}s Casas}, {Federici}, {Fedorets},
  {Fernique}, {Figueras}, {Filippi}, {Findeisen}, {Fonti}, {Fraile}, {Fraser},
  {Fr{\'e}zouls}, {Gai}, {Galleti}, {Garabato}, {Garc{\'\i}a-Sedano},
  {Garofalo}, {Garralda}, {Gavel}, {Gavras}, {Gerssen}, {Geyer}, {Giacobbe},
  {Gilmore}, {Girona}, {Giuffrida}, {Glass}, {Gomes}, {Granvik}, {Gueguen},
  {Guerrier}, {Guiraud}, {Guti{\'e}}, {Haigron}, {Hatzidimitriou}, {Hauser},
  {Haywood}, {Heiter}, {Helmi}, {Heu}, {Hilger}, {Hobbs}, {Hofmann}, {Holland},
  {Huckle}, {Hypki}, {Icardi}, {Jan{\ss}en}, {Jevardat de Fombelle}, {Jonker},
  {Juh{\'a}sz}, {Julbe}, {Karampelas}, {Kewley}, {Klar}, {Kochoska}, {Kohley},
  {Kolenberg}, {Kontizas}, {Kontizas}, {Koposov}, {Kordopatis},
  {Kostrzewa-Rutkowska}, {Koubsky}, {Lambert}, {Lanza}, {Lasne}, {Lavigne}, {Le
  Fustec}, {Le Poncin-Lafitte}, {Lebreton}, {Leccia}, {Leclerc},
  {Lecoeur-Taibi}, {Lenhardt}, {Leroux}, {Liao}, {Licata}, {Lindstr{\o}m},
  {Lister}, {Livanou}, {Lobel}, {L{\'o}pez}, {Managau}, {Mann}, {Mantelet},
  {Marchal}, {Marchant}, {Marconi}, {Marinoni}, {Marschalk{\'o}}, {Marshall},
  {Martino}, {Marton}, {Mary}, {Massari}, {Matijevi{\v{c}}}, {Mazeh},
  {McMillan}, {Messina}, {Michalik}, {Millar}, {Molina}, {Molinaro},
  {Moln{\'a}r}, {Montegriffo}, {Mor}, {Morbidelli}, {Morel}, {Morris},
  {Mulone}, {Muraveva}, {Musella}, {Nelemans}, {Nicastro}, {Noval},
  {O'Mullane}, {Ord{\'e}novic}, {Ord{\'o}{\~n}ez-Blanco}, {Osborne}, {Pagani},
  {Pagano}, {Pailler}, {Palacin}, {Palaversa}, {Panahi}, {Pawlak},
  {Piersimoni}, {Pineau}, {Plachy}, {Plum}, {Poggio}, {Poujoulet},
  {Pr{\v{s}}a}, {Pulone}, {Racero}, {Ragaini}, {Rambaux}, {Ramos-Lerate},
  {Regibo}, {Reyl{\'e}}, {Riclet}, {Ripepi}, {Riva}, {Rivard}, {Rixon},
  {Roegiers}, {Roelens}, {Romero-G{\'o}mez}, {Rowell}, {Royer}, {Ruiz-Dern},
  {Sadowski}, {Sagrist{\`a} Sell{\'e}s}, {Sahlmann}, {Salgado}, {Salguero},
  {Sanna}, {Santana- Ros}, {Sarasso}, {Savietto}, {Schultheis}, {Sciacca},
  {Segol}, {Segovia}, {S{\'e}gransan}, {Shih}, {Siltala}, {Silva}, {Smart},
  {Smith}, {Solano}, {Solitro}, {Sordo}, {Soria Nieto}, {Souchay}, {Spagna},
  {Spoto}, {Stampa}, {Steele}, {Steidelm{\"u}ller}, {Stephenson}, {Stoev},
  {Suess}, {Surdej}, {Szabados}, {Szegedi-Elek}, {Tapiador}, {Taris}, {Tauran},
  {Taylor}, {Teixeira}, {Terrett}, {Teyssandier}, {Thuillot}, {Titarenko},
  {Torra Clotet}, {Turon}, {Ulla}, {Utrilla}, {Uzzi}, {Vaillant}, {Valentini},
  {Valette}, {van Elteren}, {Van Hemelryck}, {Vaschetto}, {Vecchiato},
  {Veljanoski}, {Viala}, {Vicente}, {Vogt}, {von Essen}, {Voss}, {Votruba},
  {Voutsinas}, {Walmsley}, {Weiler}, {Wertz}, {Wevers}, {Wyrzykowski},
  {Yoldas}, {{\v{Z}}erjal}, {Ziaeepour}, {Zorec}, {Zschocke}, {Zucker},
  {Zurbach}, \& {Zwitter}}]{hrd}
{Gaia Collaboration}, {Babusiaux}, C., {van Leeuwen}, F., {et~al.}
  2018{\natexlab{a}}, \aap, 616, A10

\bibitem[{{Gaia Collaboration} {et~al.}(2018{\natexlab{b}}){Gaia
  Collaboration}, {Brown}, {Vallenari}, {Prusti}, {de Bruijne}, {Babusiaux},
  {Bailer-Jones}, {Biermann}, {Evans}, {Eyer}, {Jansen}, {Jordi}, {Klioner},
  {Lammers}, {Lindegren}, {Luri}, {Mignard}, {Panem}, {Pourbaix}, {Randich},
  {Sartoretti}, {Siddiqui}, {Soubiran}, {van Leeuwen}, {Walton}, {Arenou},
  {Bastian}, {Cropper}, {Drimmel}, {Katz}, {Lattanzi}, {Bakker}, {Cacciari},
  {Casta{\~n}eda}, {Chaoul}, {Cheek}, {De Angeli}, {Fabricius}, {Guerra},
  {Holl}, {Masana}, {Messineo}, {Mowlavi}, {Nienartowicz}, {Panuzzo},
  {Portell}, {Riello}, {Seabroke}, {Tanga}, {Th{\'e}venin}, {Gracia-Abril},
  {Comoretto}, {Garcia-Reinaldos}, {Teyssier}, {Altmann}, {Andrae}, {Audard},
  {Bellas-Velidis}, {Benson}, {Berthier}, {Blomme}, {Burgess}, {Busso},
  {Carry}, {Cellino}, {Clementini}, {Clotet}, {Creevey}, {Davidson}, {De
  Ridder}, {Delchambre}, {Dell'Oro}, {Ducourant}, {Fern{\'a}ndez-
  Hern{\'a}ndez}, {Fouesneau}, {Fr{\'e}mat}, {Galluccio}, {Garc{\'\i}a-Torres},
  {Gonz{\'a}lez-N{\'u}{\~n}ez}, {Gonz{\'a}lez-Vidal}, {Gosset}, {Guy},
  {Halbwachs}, {Hambly}, {Harrison}, {Hern{\'a}ndez}, {Hestroffer}, {Hodgkin},
  {Hutton}, {Jasniewicz}, {Jean-Antoine-Piccolo}, {Jordan}, {Korn},
  {Krone-Martins}, {Lanzafame}, {Lebzelter}, {L{\"o}ffler}, {Manteiga},
  {Marrese}, {Mart{\'\i}n-Fleitas}, {Moitinho}, {Mora}, {Muinonen}, {Osinde},
  {Pancino}, {Pauwels}, {Petit}, {Recio-Blanco}, {Richards}, {Rimoldini},
  {Robin}, {Sarro}, {Siopis}, {Smith}, {Sozzetti}, {S{\"u}veges}, {Torra}, {van
  Reeven}, {Abbas}, {Abreu Aramburu}, {Accart}, {Aerts}, {Altavilla},
  {{\'A}lvarez}, {Alvarez}, {Alves}, {Anderson}, {Andrei}, {Anglada Varela},
  {Antiche}, {Antoja}, {Arcay}, {Astraatmadja}, {Bach}, {Baker},
  {Balaguer-N{\'u}{\~n}ez}, {Balm}, {Barache}, {Barata}, {Barbato}, {Barblan},
  {Barklem}, {Barrado}, {Barros}, {Barstow}, {Bartholom{\'e} Mu{\~n}oz},
  {Bassilana}, {Becciani}, {Bellazzini}, {Berihuete}, {Bertone}, {Bianchi},
  {Bienaym{\'e}}, {Blanco-Cuaresma}, {Boch}, {Boeche}, {Bombrun}, {Borrachero},
  {Bossini}, {Bouquillon}, {Bourda}, {Bragaglia}, {Bramante}, {Breddels},
  {Bressan}, {Brouillet}, {Br{\"u}semeister}, {Brugaletta}, {Bucciarelli},
  {Burlacu}, {Busonero}, {Butkevich}, {Buzzi}, {Caffau}, {Cancelliere},
  {Cannizzaro}, {Cantat-Gaudin}, {Carballo}, {Carlucci}, {Carrasco},
  {Casamiquela}, {Castellani}, {Castro-Ginard}, {Charlot}, {Chemin},
  {Chiavassa}, {Cocozza}, {Costigan}, {Cowell}, {Crifo}, {Crosta}, {Crowley},
  {Cuypers}, {Dafonte}, {Damerdji}, {Dapergolas}, {David}, {David}, {de
  Laverny}, {De Luise}, {De March}, {de Martino}, {de Souza}, {de Torres},
  {Debosscher}, {del Pozo}, {Delbo}, {Delgado}, {Delgado}, {Di Matteo},
  {Diakite}, {Diener}, {Distefano}, {Dolding}, {Drazinos}, {Dur{\'a}n},
  {Edvardsson}, {Enke}, {Eriksson}, {Esquej}, {Eynard Bontemps}, {Fabre},
  {Fabrizio}, {Faigler}, {Falc{\~a}o}, {Farr{\`a}s Casas}, {Federici},
  {Fedorets}, {Fernique}, {Figueras}, {Filippi}, {Findeisen}, {Fonti},
  {Fraile}, {Fraser}, {Fr{\'e}zouls}, {Gai}, {Galleti}, {Garabato},
  {Garc{\'\i}a-Sedano}, {Garofalo}, {Garralda}, {Gavel}, {Gavras}, {Gerssen},
  {Geyer}, {Giacobbe}, {Gilmore}, {Girona}, {Giuffrida}, {Glass}, {Gomes},
  {Granvik}, {Gueguen}, {Guerrier}, {Guiraud}, {Guti{\'e}rrez-S{\'a}nchez},
  {Haigron}, {Hatzidimitriou}, {Hauser}, {Haywood}, {Heiter}, {Helmi}, {Heu},
  {Hilger}, {Hobbs}, {Hofmann}, {Holland}, {Huckle}, {Hypki}, {Icardi},
  {Jan{\ss}en}, {Jevardat de Fombelle}, {Jonker}, {Juh{\'a}sz}, {Julbe},
  {Karampelas}, {Kewley}, {Klar}, {Kochoska}, {Kohley}, {Kolenberg},
  {Kontizas}, {Kontizas}, {Koposov}, {Kordopatis}, {Kostrzewa-Rutkowska},
  {Koubsky}, {Lambert}, {Lanza}, {Lasne}, {Lavigne}, {Le Fustec}, {Le
  Poncin-Lafitte}, {Lebreton}, {Leccia}, {Leclerc}, {Lecoeur-Taibi},
  {Lenhardt}, {Leroux}, {Liao}, {Licata}, {Lindstr{\o}m}, {Lister}, {Livanou},
  {Lobel}, {L{\'o}pez}, {Managau}, {Mann}, {Mantelet}, {Marchal}, {Marchant},
  {Marconi}, {Marinoni}, {Marschalk{\'o}}, {Marshall}, {Martino}, {Marton},
  {Mary}, {Massari}, {Matijevi{\v{c}}}, {Mazeh}, {McMillan}, {Messina},
  {Michalik}, {Millar}, {Molina}, {Molinaro}, {Moln{\'a}r}, {Montegriffo},
  {Mor}, {Morbidelli}, {Morel}, {Morris}, {Mulone}, {Muraveva}, {Musella},
  {Nelemans}, {Nicastro}, {Noval}, {O'Mullane}, {Ord{\'e}novic},
  {Ord{\'o}{\~n}ez-Blanco}, {Osborne}, {Pagani}, {Pagano}, {Pailler},
  {Palacin}, {Palaversa}, {Panahi}, {Pawlak}, {Piersimoni}, {Pineau}, {Plachy},
  {Plum}, {Poggio}, {Poujoulet}, {Pr{\v{s}}a}, {Pulone}, {Racero}, {Ragaini},
  {Rambaux}, {Ramos-Lerate}, {Regibo}, {Reyl{\'e}}, {Riclet}, {Ripepi}, {Riva},
  {Rivard}, {Rixon}, {Roegiers}, {Roelens}, {Romero-G{\'o}mez}, {Rowell},
  {Royer}, {Ruiz-Dern}, {Sadowski}, {Sagrist{\`a} Sell{\'e}s}, {Sahlmann},
  {Salgado}, {Salguero}, {Sanna}, {Santana- Ros}, {Sarasso}, {Savietto},
  {Schultheis}, {Sciacca}, {Segol}, {Segovia}, {S{\'e}gransan}, {Shih},
  {Siltala}, {Silva}, {Smart}, {Smith}, {Solano}, {Solitro}, {Sordo}, {Soria
  Nieto}, {Souchay}, {Spagna}, {Spoto}, {Stampa}, {Steele},
  {Steidelm{\"u}ller}, {Stephenson}, {Stoev}, {Suess}, {Surdej}, {Szabados},
  {Szegedi-Elek}, {Tapiador}, {Taris}, {Tauran}, {Taylor}, {Teixeira},
  {Terrett}, {Teyssandier}, {Thuillot}, {Titarenko}, {Torra Clotet}, {Turon},
  {Ulla}, {Utrilla}, {Uzzi}, {Vaillant}, {Valentini}, {Valette}, {van Elteren},
  {Van Hemelryck}, {van Leeuwen}, {Vaschetto}, {Vecchiato}, {Veljanoski},
  {Viala}, {Vicente}, {Vogt}, {von Essen}, {Voss}, {Votruba}, {Voutsinas},
  {Walmsley}, {Weiler}, {Wertz}, {Wevers}, {Wyrzykowski}, {Yoldas},
  {{\v{Z}}erjal}, {Ziaeepour}, {Zorec}, {Zschocke}, {Zucker}, {Zurbach}, \&
  {Zwitter}}]{gaia_dr2}
{Gaia Collaboration}, {Brown}, A.~G.~A., {Vallenari}, A., {et~al.}
  2018{\natexlab{b}}, \aap, 616, A1

\bibitem[{{Gaia Collaboration} {et~al.}(2016){Gaia Collaboration}, {Prusti},
  {de Bruijne}, {Brown}, {Vallenari}, {Babusiaux}, {Bailer-Jones}, {Bastian},
  {Biermann}, {Evans}, {Eyer}, {Jansen}, {Jordi}, {Klioner}, {Lammers},
  {Lindegren}, {Luri}, {Mignard}, {Milligan}, {Panem}, {Poinsignon},
  {Pourbaix}, {Randich}, {Sarri}, {Sartoretti}, {Siddiqui}, {Soubiran},
  {Valette}, {van Leeuwen}, {Walton}, {Aerts}, {Arenou}, {Cropper}, {Drimmel},
  {H{\o}g}, {Katz}, {Lattanzi}, {O'Mullane}, {Grebel}, {Holland}, {Huc},
  {Passot}, {Bramante}, {Cacciari}, {Casta{\~n}eda}, {Chaoul}, {Cheek}, {De
  Angeli}, {Fabricius}, {Guerra}, {Hern{\'a}ndez}, {Jean-Antoine-Piccolo},
  {Masana}, {Messineo}, {Mowlavi}, {Nienartowicz}, {Ord{\'o}{\~n}ez- Blanco},
  {Panuzzo}, {Portell}, {Richards}, {Riello}, {Seabroke}, {Tanga},
  {Th{\'e}venin}, {Torra}, {Els}, {Gracia- Abril}, {Comoretto},
  {Garcia-Reinaldos}, {Lock}, {Mercier}, {Altmann}, {Andrae}, {Astraatmadja},
  {Bellas-Velidis}, {Benson}, {Berthier}, {Blomme}, {Busso}, {Carry},
  {Cellino}, {Clementini}, {Cowell}, {Creevey}, {Cuypers}, {Davidson}, {De
  Ridder}, {de Torres}, {Delchambre}, {Dell'Oro}, {Ducourant}, {Fr{\'e}mat},
  {Garc{\'\i}a-Torres}, {Gosset}, {Halbwachs}, {Hambly}, {Harrison}, {Hauser},
  {Hestroffer}, {Hodgkin}, {Huckle}, {Hutton}, {Jasniewicz}, {Jordan},
  {Kontizas}, {Korn}, {Lanzafame}, {Manteiga}, {Moitinho}, {Muinonen},
  {Osinde}, {Pancino}, {Pauwels}, {Petit}, {Recio-Blanco}, {Robin}, {Sarro},
  {Siopis}, {Smith}, {Smith}, {Sozzetti}, {Thuillot}, {van Reeven}, {Viala},
  {Abbas}, {Abreu Aramburu}, {Accart}, {Aguado}, {Allan}, {Allasia},
  {Altavilla}, {{\'A}lvarez}, {Alves}, {Anderson}, {Andrei}, {Anglada Varela},
  {Antiche}, {Antoja}, {Ant{\'o}n}, {Arcay}, {Atzei}, {Ayache}, {Bach},
  {Baker}, {Balaguer-N{\'u}{\~n}ez}, {Barache}, {Barata}, {Barbier}, {Barblan},
  {Baroni}, {Barrado y Navascu{\'e}s}, {Barros}, {Barstow}, {Becciani},
  {Bellazzini}, {Bellei}, {Bello Garc{\'\i}a}, {Belokurov}, {Bendjoya},
  {Berihuete}, {Bianchi}, {Bienaym{\'e}}, {Billebaud}, {Blagorodnova},
  {Blanco-Cuaresma}, {Boch}, {Bombrun}, {Borrachero}, {Bouquillon}, {Bourda},
  {Bouy}, {Bragaglia}, {Breddels}, {Brouillet}, {Br{\"u}semeister},
  {Bucciarelli}, {Budnik}, {Burgess}, {Burgon}, {Burlacu}, {Busonero}, {Buzzi},
  {Caffau}, {Cambras}, {Campbell}, {Cancelliere}, {Cantat-Gaudin}, {Carlucci},
  {Carrasco}, {Castellani}, {Charlot}, {Charnas}, {Charvet}, {Chassat},
  {Chiavassa}, {Clotet}, {Cocozza}, {Collins}, {Collins}, {Costigan}, {Crifo},
  {Cross}, {Crosta}, {Crowley}, {Dafonte}, {Damerdji}, {Dapergolas}, {David},
  {David}, {De Cat}, {de Felice}, {de Laverny}, {De Luise}, {De March}, {de
  Martino}, {de Souza}, {Debosscher}, {del Pozo}, {Delbo}, {Delgado},
  {Delgado}, {di Marco}, {Di Matteo}, {Diakite}, {Distefano}, {Dolding}, {Dos
  Anjos}, {Drazinos}, {Dur{\'a}n}, {Dzigan}, {Ecale}, {Edvardsson}, {Enke},
  {Erdmann}, {Escolar}, {Espina}, {Evans}, {Eynard Bontemps}, {Fabre},
  {Fabrizio}, {Faigler}, {Falc{\~a}o}, {Farr{\`a}s Casas}, {Faye}, {Federici},
  {Fedorets}, {Fern{\'a}ndez-Hern{\'a}ndez}, {Fernique}, {Fienga}, {Figueras},
  {Filippi}, {Findeisen}, {Fonti}, {Fouesneau}, {Fraile}, {Fraser}, {Fuchs},
  {Furnell}, {Gai}, {Galleti}, {Galluccio}, {Garabato}, {Garc{\'\i}a-Sedano},
  {Gar{\'e}}, {Garofalo}, {Garralda}, {Gavras}, {Gerssen}, {Geyer}, {Gilmore},
  {Girona}, {Giuffrida}, {Gomes}, {Gonz{\'a}lez-Marcos},
  {Gonz{\'a}lez-N{\'u}{\~n}ez}, {Gonz{\'a}lez-Vidal}, {Granvik}, {Guerrier},
  {Guillout}, {Guiraud}, {G{\'u}rpide}, {Guti{\'e}rrez-S{\'a}nchez}, {Guy},
  {Haigron}, {Hatzidimitriou}, {Haywood}, {Heiter}, {Helmi}, {Hobbs},
  {Hofmann}, {Holl}, {Holland}, {Hunt}, {Hypki}, {Icardi}, {Irwin}, {Jevardat
  de Fombelle}, {Jofr{\'e}}, {Jonker}, {Jorissen}, {Julbe}, {Karampelas},
  {Kochoska}, {Kohley}, {Kolenberg}, {Kontizas}, {Koposov}, {Kordopatis},
  {Koubsky}, {Kowalczyk}, {Krone-Martins}, {Kudryashova}, {Kull}, {Bachchan},
  {Lacoste-Seris}, {Lanza}, {Lavigne}, {Le Poncin-Lafitte}, {Lebreton},
  {Lebzelter}, {Leccia}, {Leclerc}, {Lecoeur-Taibi}, {Lemaitre}, {Lenhardt},
  {Leroux}, {Liao}, {Licata}, {Lindstr{\o}m}, {Lister}, {Livanou}, {Lobel},
  {L{\"o}ffler}, {L{\'o}pez}, {Lopez-Lozano}, {Lorenz}, {Loureiro},
  {MacDonald}, {Magalh{\~a}es Fernandes}, {Managau}, {Mann}, {Mantelet},
  {Marchal}, {Marchant}, {Marconi}, {Marie}, {Marinoni}, {Marrese},
  {Marschalk{\'o}}, {Marshall}, {Mart{\'\i}n-Fleitas}, {Martino}, {Mary},
  {Matijevi{\v{c}}}, {Mazeh}, {McMillan}, {Messina}, {Mestre}, {Michalik},
  {Millar}, {Miranda}, {Molina}, {Molinaro}, {Molinaro}, {Moln{\'a}r},
  {Moniez}, {Montegriffo}, {Monteiro}, {Mor}, {Mora}, {Morbidelli}, {Morel},
  {Morgenthaler}, {Morley}, {Morris}, {Mulone}, {Muraveva}, {Musella},
  {Narbonne}, {Nelemans}, {Nicastro}, {Noval}, {Ord{\'e}novic},
  {Ordieres-Mer{\'e}}, {Osborne}, {Pagani}, {Pagano}, {Pailler}, {Palacin},
  {Palaversa}, {Parsons}, {Paulsen}, {Pecoraro}, {Pedrosa}, {Pentik{\"a}inen},
  {Pereira}, {Pichon}, {Piersimoni}, {Pineau}, {Plachy}, {Plum}, {Poujoulet},
  {Pr{\v{s}}a}, {Pulone}, {Ragaini}, {Rago}, {Rambaux}, {Ramos-Lerate},
  {Ranalli}, {Rauw}, {Read}, {Regibo}, {Renk}, {Reyl{\'e}}, {Ribeiro},
  {Rimoldini}, {Ripepi}, {Riva}, {Rixon}, {Roelens}, {Romero-G{\'o}mez},
  {Rowell}, {Royer}, {Rudolph}, {Ruiz-Dern}, {Sadowski}, {Sagrist{\`a}
  Sell{\'e}s}, {Sahlmann}, {Salgado}, {Salguero}, {Sarasso}, {Savietto},
  {Schnorhk}, {Schultheis}, {Sciacca}, {Segol}, {Segovia}, {Segransan},
  {Serpell}, {Shih}, {Smareglia}, {Smart}, {Smith}, {Solano}, {Solitro},
  {Sordo}, {Soria Nieto}, {Souchay}, {Spagna}, {Spoto}, {Stampa}, {Steele},
  {Steidelm{\"u}ller}, {Stephenson}, {Stoev}, {Suess}, {S{\"u}veges}, {Surdej},
  {Szabados}, {Szegedi-Elek}, {Tapiador}, {Taris}, {Tauran}, {Taylor},
  {Teixeira}, {Terrett}, {Tingley}, {Trager}, {Turon}, {Ulla}, {Utrilla},
  {Valentini}, {van Elteren}, {Van Hemelryck}, {van Leeuwen}, {Varadi},
  {Vecchiato}, {Veljanoski}, {Via}, {Vicente}, {Vogt}, {Voss}, {Votruba},
  {Voutsinas}, {Walmsley}, {Weiler}, {Weingrill}, {Werner}, {Wevers},
  {Whitehead}, {Wyrzykowski}, {Yoldas}, {{\v{Z}}erjal}, {Zucker}, {Zurbach},
  {Zwitter}, {Alecu}, {Allen}, {Allende Prieto}, {Amorim},
  {Anglada-Escud{\'e}}, {Arsenijevic}, {Azaz}, {Balm}, {Beck}, {Bernstein},
  {Bigot}, {Bijaoui}, {Blasco}, {Bonfigli}, {Bono}, {Boudreault}, {Bressan},
  {Brown}, {Brunet}, {Bunclark}, {Buonanno}, {Butkevich}, {Carret}, {Carrion},
  {Chemin}, {Ch{\'e}reau}, {Corcione}, {Darmigny}, {de Boer}, {de Teodoro}, {de
  Zeeuw}, {Delle Luche}, {Domingues}, {Dubath}, {Fodor}, {Fr{\'e}zouls},
  {Fries}, {Fustes}, {Fyfe}, {Gallardo}, {Gallegos}, {Gardiol}, {Gebran},
  {Gomboc}, {G{\'o}mez}, {Grux}, {Gueguen}, {Heyrovsky}, {Hoar}, {Iannicola},
  {Isasi Parache}, {Janotto}, {Joliet}, {Jonckheere}, {Keil}, {Kim},
  {Klagyivik}, {Klar}, {Knude}, {Kochukhov}, {Kolka}, {Kos}, {Kutka}, {Lainey},
  {LeBouquin}, {Liu}, {Loreggia}, {Makarov}, {Marseille}, {Martayan},
  {Martinez-Rubi}, {Massart}, {Meynadier}, {Mignot}, {Munari}, {Nguyen},
  {Nordlander}, {Ocvirk}, {O'Flaherty}, {Olias Sanz}, {Ortiz}, {Osorio},
  {Oszkiewicz}, {Ouzounis}, {Palmer}, {Park}, {Pasquato}, {Peltzer}, {Peralta},
  {P{\'e}turaud}, {Pieniluoma}, {Pigozzi}, {Poels}, {Prat}, {Prod'homme},
  {Raison}, {Rebordao}, {Risquez}, {Rocca-Volmerange}, {Rosen}, {Ruiz-Fuertes},
  {Russo}, {Sembay}, {Serraller Vizcaino}, {Short}, {Siebert}, {Silva},
  {Sinachopoulos}, {Slezak}, {Soffel}, {Sosnowska}, {Strai{\v{z}}ys}, {ter
  Linden}, {Terrell}, {Theil}, {Tiede}, {Troisi}, {Tsalmantza}, {Tur},
  {Vaccari}, {Vachier}, {Valles}, {Van Hamme}, {Veltz}, {Virtanen}, {Wallut},
  {Wichmann}, {Wilkinson}, {Ziaeepour}, \& {Zschocke}}]{gaia_mission}
{Gaia Collaboration}, {Prusti}, T., {de Bruijne}, J.~H.~J., {et~al.} 2016,
  \aap, 595, A1

\bibitem[{Hunter(2007)}]{matplotlib}
Hunter, J.~D. 2007, Computing In Science \& Engineering, 9, 90

\bibitem[{{Ibata} {et~al.}(2019){Ibata}, {Malhan}, \& {Martin}}]{ibata}
{Ibata}, R., {Malhan}, K., \& {Martin}, N. 2019, arXiv e-prints,
  arXiv:1901.07566

\bibitem[{Jones {et~al.}(2001)Jones, Oliphant, Peterson, {et~al.}}]{scipy}
Jones, E., Oliphant, T., Peterson, P., {et~al.} 2001, {SciPy}: Open source
  scientific tools for {Python}

\bibitem[{{Kraus} \& {Hillenbrand}(2007)}]{kraus}
{Kraus}, A.~L. \& {Hillenbrand}, L.~A. 2007, \aj, 134, 2340

\bibitem[{{Kroupa}(2001)}]{kroupa01}
{Kroupa}, P. 2001, \mnras, 322, 231

\bibitem[{{Meingast} \& {Alves}(2019)}]{meingast18}
{Meingast}, S. \& {Alves}, J. 2019, \aap, 621, L3

\bibitem[{{Meingast} {et~al.}(2019){Meingast}, {Alves}, \&
  {F{\"u}rnkranz}}]{meingast19}
{Meingast}, S., {Alves}, J., \& {F{\"u}rnkranz}, V. 2019, arXiv e-prints,
  arXiv:1901.06387

\bibitem[{{Netopil} {et~al.}(2016){Netopil}, {Paunzen}, {Heiter}, \&
  {Soubiran}}]{netopil}
{Netopil}, M., {Paunzen}, E., {Heiter}, U., \& {Soubiran}, C. 2016, \aap, 585,
  A150

\bibitem[{{Paunzen}(2008)}]{webda}
{Paunzen}, E. 2008, Contributions of the Astronomical Observatory Skalnate
  Pleso, 38, 435

\bibitem[{Pedregosa {et~al.}(2011)Pedregosa, Varoquaux, Gramfort, Michel,
  Thirion, Grisel, Blondel, Prettenhofer, Weiss, Dubourg, Vanderplas, Passos,
  Cournapeau, Brucher, Perrot, \& Duchesnay}]{scikit-learn}
Pedregosa, F., Varoquaux, G., Gramfort, A., {et~al.} 2011, Journal of Machine
  Learning Research, 12, 2825

\bibitem[{{Riedel} {et~al.}(2017){Riedel}, {Blunt}, {Lambrides}, {Rice},
  {Cruz}, \& {Faherty}}]{riedel}
{Riedel}, A.~R., {Blunt}, S.~C., {Lambrides}, E.~L., {et~al.} 2017, \aj, 153,
  95

\bibitem[{{R{\"o}ser} {et~al.}(2019){R{\"o}ser}, {Schilbach}, \&
  {Goldman}}]{roser}
{R{\"o}ser}, S., {Schilbach}, E., \& {Goldman}, B. 2019, \aap, 621, L2

\bibitem[{{Scheepmaker} {et~al.}(2009){Scheepmaker}, {Lamers}, {Anders}, \&
  {Larsen}}]{scheepmaker}
{Scheepmaker}, R.~A., {Lamers}, H.~J.~G.~L.~M., {Anders}, P., \& {Larsen},
  S.~S. 2009, \aap, 494, 81

\bibitem[{{Tang} {et~al.}(2018){Tang}, {Chen}, {Chiang}, {Jose}, {Herczeg}, \&
  {Goldman}}]{tang}
{Tang}, S.-Y., {Chen}, W.~P., {Chiang}, P.~S., {et~al.} 2018, \apj, 862, 106

\bibitem[{{van der Walt} {et~al.}(2011){van der Walt}, {Colbert}, \&
  {Varoquaux}}]{numpy}
{van der Walt}, S., {Colbert}, S.~C., \& {Varoquaux}, G. 2011, Computing in
  Science and Engg., 13, 22

\bibitem[{van~der Walt {et~al.}(2014)van~der Walt, {S}ch\"onberger,
  {Nunez-Iglesias}, {B}oulogne, {W}arner, {Y}ager, {G}ouillart, {Y}u, \& the
  scikit-image contributors}]{scikit-image}
van~der Walt, S., {S}ch\"onberger, J.~L., {Nunez-Iglesias}, J., {et~al.} 2014,
  PeerJ, 2, e453

\bibitem[{{Wenger} {et~al.}(2000){Wenger}, {Ochsenbein}, {Egret}, {Dubois},
  {Bonnarel}, {Borde}, {Genova}, {Jasniewicz}, {Lalo{\"e}}, {Lesteven}, \&
  {Monier}}]{simbad}
{Wenger}, M., {Ochsenbein}, F., {Egret}, D., {et~al.} 2000, \aaps, 143, 9

\end{thebibliography}
\clearpage

\begin{appendix}

\section{Supplementary plots and tables}
\label{sec:app}

Here we provide supplementary material. Table~\ref{tab:table} shows fundamental properties of both groups. The parameters are average values obtained from our final source selection. The numbers in parenthesis correspond to the group members with radial-velocity measurements. As a measure of the dispersion in each parameter, we additionally quote the standard deviation of the obtained values. The comparably high dispersion of the vertical velocity component of \mel can be explained by outliers in our selection and the strong dependence on radial-velocity measurements. Our determined values for \mel are in excellent agreement with the literature. For example, \citet{tang} estimated an age of $\sim$ 800 Myr and a distance of $\sim$ \SI{86.7}{pc}. \citet{riedel} published position and velocity coordinates of $(X,Y,Z) = (-6.706,-6.308,87.522)$ \SI{}{pc} and $(U,V,W) = (-2.512, -5.417,-1.204)$ \SI{}{\km \per \second}, and \citet{kraus} determined the mean cluster proper motion as $(\mu_{\alpha},\mu_{\delta}) = (-11.5,-9.5)$ mas yr$^{-1}$. 

\begin{figure} [htp]
        \centering
        \resizebox{1.0\hsize}{!}{\includegraphics[]{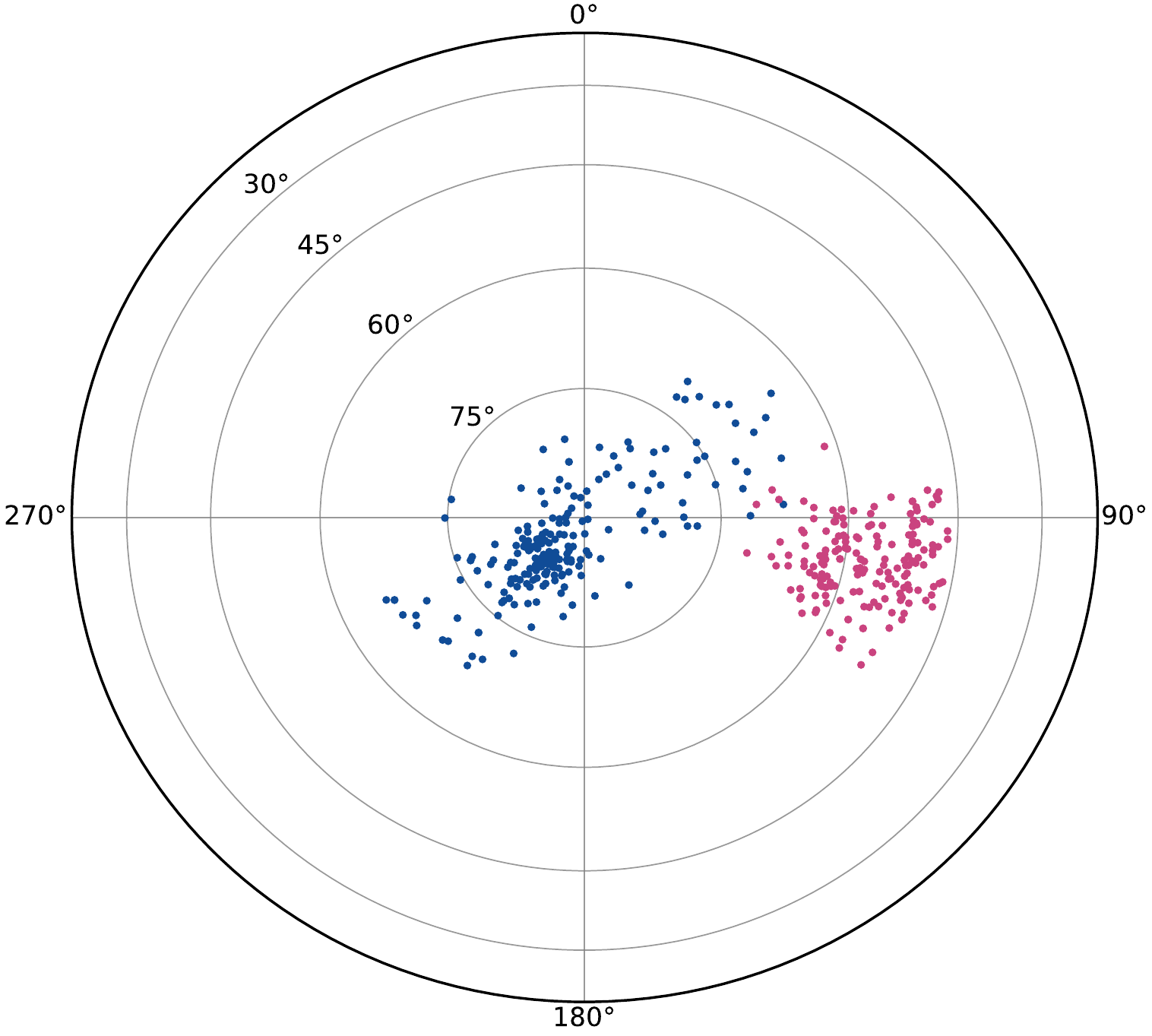}}
        \caption[]{Distribution of our final member selection displayed in a slant orthographic projection with the north Galactic pole at its center.}
    \label{img:proj}
\end{figure}

\begin{figure} [hbp]
        \centering
        \resizebox{1.0\hsize}{!}{\includegraphics[]{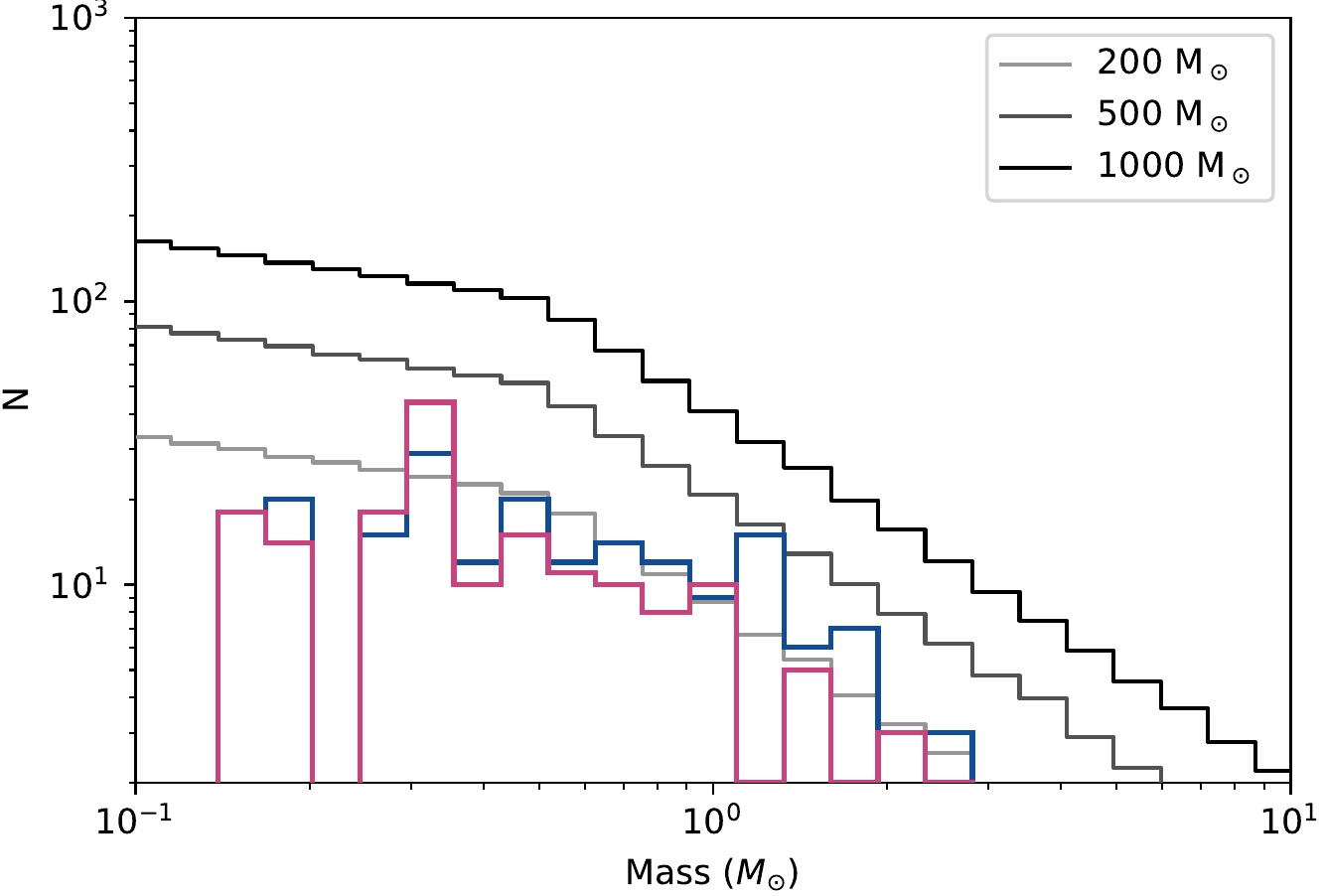}}
        \caption[]{Mass functions of all member sources for both groups, with a series of IMFs on top.}
    \label{img:mass}
\end{figure}

Figure~\ref{img:proj} shows the distribution of our member selection in a slant orthographic projection, centered at the north Galactic pole. Figure~\ref{img:mass} displays the mass distribution of both groups, as well as a series of IMFs. As discussed in Sect.~\ref{sec:structure}, both groups match well with a \SI{200}{M_\odot} IMF. Due to incompleteness however, we estimated their birth masses to be closer to \SI{500}{M_\odot}, in contrast to the  $\sim$\SI{100}{M_\odot} measured for the present-day mass of the cluster core \citep{casewell, kraus}.

\begin{figure*} 
        \centering
        \resizebox{1.0\hsize}{!}{\includegraphics[]{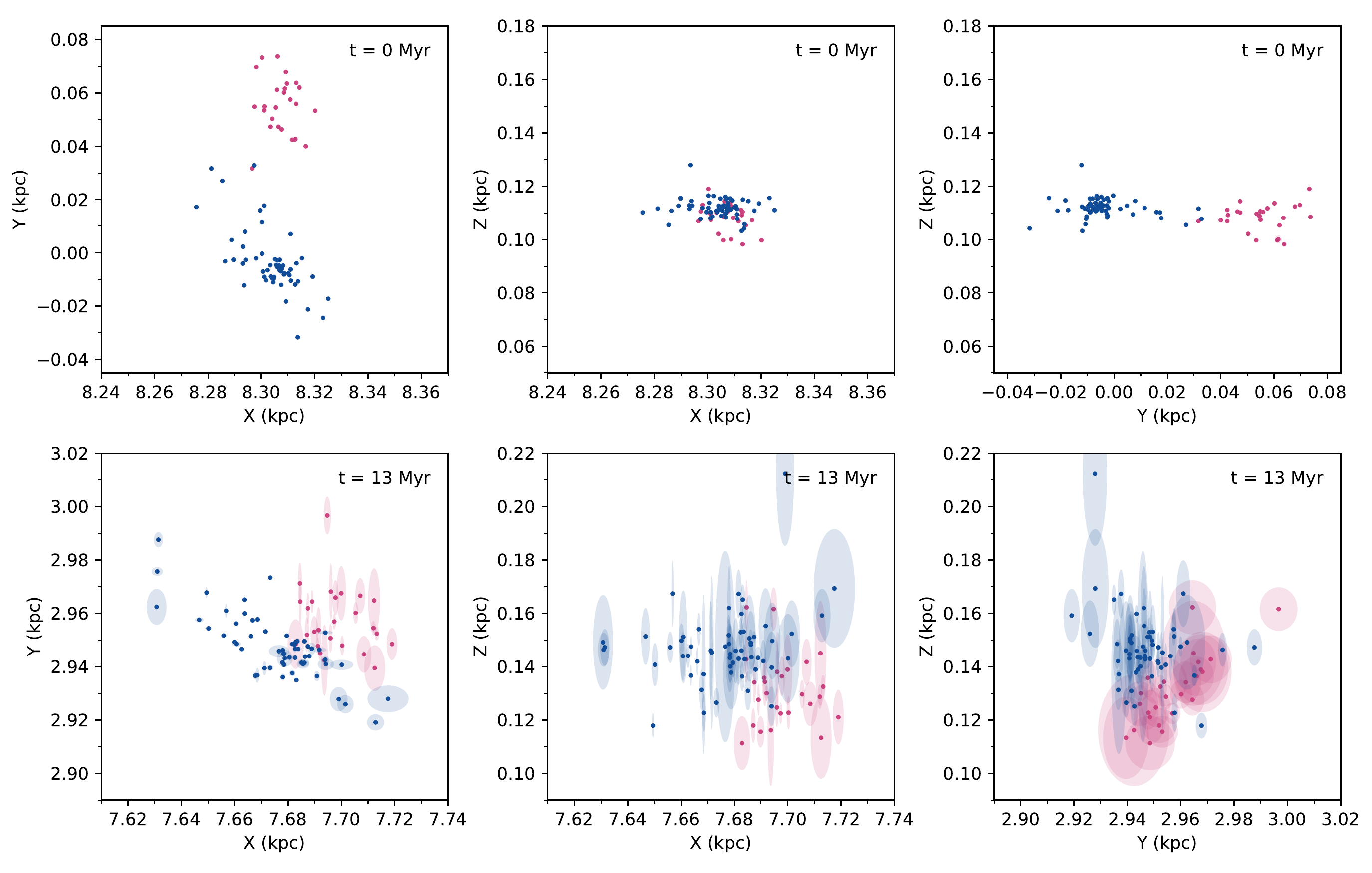}}
        \caption[]{The top row illustrates the current position of all selected stars with radial-velocity measurements ($\sigma_{rv} < 2$ km s$^{-1}$) in the Galactocentric Coordinate frame. The bottom row shows the positions of these sources at t = 13 Myr. The ellipses represent the errors of orbit integration (3-$\sigma$). The large errors in the direction of Z correspond to the large errors in radial velocities and are responsible for the disappearance of the flatness of the structure parallel to the Galactic plane.}
    \label{img:galpy}
\end{figure*}

Figure~\ref{img:galpy} illustrates the positions of the member stars of \mel and the new group both now and at the time of minimum distance in Galactocentric Cartesian coordinates. The mean position values are represented as dots, and the 3-$\sigma$ errors are illustrated with transparent ellipses. As described in Sect.~\ref{sec:structure}, both groups appear as flattened structures parallel to the Galactic plane. However, we do not find a similarly flat arrangement in \SI{13}{Myr}. This is however caused mostly by measurement errors associated with the radial velocities. 

\begin{table} 
    \begin{tabular*}{\linewidth}{l @{\extracolsep{\fill}} c c}
        \hline\hline
        Property &  \mel & New group \\
    \hline
    Candidate members            & 214 (67)& 177 (31) \\
    Age estimate (Myr)           & 700 & 400  \\
    Ra (deg)            & 189.28  & 215.82 \\
    Dec (deg)            & 26.24  & 55.20  \\
    X (pc)            & -4.01 $\pm$ 9.62 & -8.34 $\pm$ 7.16 \\
    Y  (pc)          & -3.48 $\pm$ 12.60 & 52.97 $\pm$ 12.24 \\
    Z  (pc)          & 85.05 $\pm$ 3.28 & 82.17 $\pm$ 4.65  \\
    d (pc)           & 86.67 $\pm$ 3.51 & 98.93 $\pm$ 7.94 \\
    rv (\SI{}{\km \per \second})           & -0.83 $\pm$ 4.78 & -6.00 $\pm$ 1.81 \\
    v$_r$ (\SI{}{\km \per \second})      & 8.91 $\pm$ 0.61 & 6.04 $\pm$ 0.48 \\
    v$_\phi$ (\SI{}{\km \per \second}) & 226.73 $\pm$ 1.18 & 223.21 $\pm$ 1.07 \\
    v$_z$ (\SI{}{\km \per \second})      & 6.06 $\pm$ 4.67 & 5.79 $\pm$ 1.25\\
    U (\SI{}{\km \per \second})          & -2.28 $\pm$ 0.71 & -3.56 $\pm$ 0.63\\
    V  (\SI{}{\km \per \second})        & -5.51 $\pm$ 1.19 & -9.08 $\pm$ 1.07 \\
    W  (\SI{}{\km \per \second})          & -1.20 $\pm$ 4.67 & -1.47 $\pm$ 1.25\\
    v$_{\alpha}$ (\SI{}{\km \per \second})   &  -4.80 $\pm$ 0.41 & -7.81 $\pm$ 0.50 \\
    v$_{\delta}$ (\SI{}{\km \per \second})   &   -3.53 $\pm$ 0.45  & -1.54 $\pm$ 0.73 \\
    $\mu_{\alpha}$ (mas yr$^{-1}$)            & -11.68 $\pm$ 1.10 & -16.75 $\pm$ 1.97\\
    $\mu_{\delta}$ (mas yr$^{-1}$)           & -8.59 $\pm$ 1.13 & -3.38  $\pm$ 1.76\\
    $\sigma _{v,\mathrm{3D}}$ (\SI{}{\km \per \second}) & 1.23  & 1.08   \\
    \hline
        \end{tabular*}
    \caption{Fundamental properties of our member selection for \mel and the new group.}
        \label{tab:table}
\end{table}

\begin{table*}
    \begin{tabular*}{\linewidth}{l @{\extracolsep{\fill}} c c c c c c c c}
        \hline\hline
\textit{Gaia} DR2 source\_id     &       Ra      &       Dec     &         X     &            Y  &       Z       &       v$_{\alpha}$   &        v$_{\delta}$         \\
        &       (deg)   &       (deg)   &       (pc)    &       (pc)   &       (pc)    &       (\SI{}{\km \per \second})  &       (\SI{}{\km \per \second})  \\
\hline                                                                                                                                  
1259389659361730048 & 214.82614 & 26.32159 & 24.11 & 17.26 & 83.21 & -3.9 & -2.82 \\
1259987931126020736 & 212.32184 & 26.65464 & 21.23 & 15.32 & 83.83 & -3.96 & -2.74 \\
1260123858250996608 & 212.78305 & 27.52297 & 20.49 & 16.48 & 82.45 & -4.12 & -2.93 \\
1260617607691437952 & 214.32504 & 28.43993 & 23.09 & 20.71 & 90.01 & -3.71 & -2.41 \\
1285098955638193792 & 215.03606 & 30.4291 & 19.93 & 22.01 & 83.01 & -4.93 & -2.76 \\
\hline
1489389418670610816 & 222.51166 & 42.15855 & 13.95 & 44.38 & 85.93 & -7.59 & -2.14 \\
1497425469984297088 & 208.26637 & 41.39739 & 2.36 & 28.91 & 83.94 & -7.96 & -2.47 \\
1498322916287022976 & 211.56849 & 41.59869 & 4.94 & 31.51 & 82.77 & -7.57 & -2.49 \\
1499294845909337344 & 211.22797 & 42.92375 & 3.13 & 31.67 & 79.9 & -7.75 & -2.51 \\
1503770755884281344 & 206.30482 & 46.31112 & -4.2 & 32.06 & 80.3 & -8.02 & -2.49 \\
\hline
    \end{tabular*}
    \caption{The top five entries show a subsample of our selected Mel 111 members, the five bottom sources belong to the new group. The full selection, including additional columns will be made available online via CDS.}
    \label{tab:sources}
\end{table*}

\end{appendix}

\end{document}